\documentclass[12pt] {article}
\usepackage{psfrag}
\usepackage{graphicx}
\usepackage{latexsym,amsfonts}
\usepackage{amssymb}
\begin{document}
\setcounter{page}{1}
\def\theequation{\arabic{section}.\arabic{equation}}
\def\theequation{\thesection.\arabic{equation}}
\setcounter{section}{0}

\title{On the vacua in the massless Thirring model}

\author{M. Faber\thanks{E--mail: faber@kph.tuwien.ac.at, Tel.:
+43--1--58801--14261, Fax: +43--1--58801--14299} ~~and~
A. N. Ivanov\thanks{E--mail: ivanov@kph.tuwien.ac.at, Tel.:
+43--1--58801--14261, Fax: +43--1--58801--14299}~\thanks{Permanent
Address: State Polytechnical University, Department of Nuclear
Physics, 195251 St. Petersburg, Russian Federation}}

\date{\today}

\maketitle

\begin{center}
{\it Atominstitut der \"Osterreichischen Universit\"aten,
Arbeitsbereich Kernphysik und Nukleare Astrophysik, Technische
Universit\"at Wien, \\ Wiedner Hauptstr. 8-10, A-1040 Wien,
\"Osterreich }
\end{center}

\begin{center}
\begin{abstract}
  We calculate the most general effective potential for the massless
  Thirring model in dependence on all local fields of collective
  fermion--antifermion excitations. We analyse the minima of this
  potential describing different vacua of the quantum system.  We
  confirm the existence of the absolute minimum found in EPJC 20, 723
  (2001) corresponding to a chirally broken phase of the massless
  Thirring model. As has been shown in EPJC 20, 723 (2001) this
  minimum is stable under quantum fluctuations.
\end{abstract}

PACS: 02.30.Ik, 11.10.Cd, 11.10.Ef, 11.10.Kk
\end{center}

\newpage

\section{Introduction}
\setcounter{equation}{0}

The massless Thirring model \cite{WT58} is a theory of
a self--coupled Dirac field $\psi(x)$
\begin{eqnarray}\label{label1.1}
{\cal L}_{\rm Th}(x) =
:\bar{\psi}(x)i\gamma^{\mu}\partial_{\mu}\psi(x):_{\mu} -
\frac{1}{2}\,g:\bar{\psi}(x)\gamma^{\mu}\psi(x)\bar{\psi}(x)
\gamma_{\mu}\psi(x):_{\mu},
\end{eqnarray}
where $g$ is a dimensionless coupling constant that can be both
positive and negative as well and $\psi(x)$ is a spinor field with two
components $\psi_1(x)$ and $\psi_2(x)$\,\footnote{The Dirac
$\gamma$--matrices are defined by \cite{FI1}: $\gamma^{\mu} =
(\gamma^0 = \sigma_1, \gamma^1 = -i\sigma_2)$ and $\gamma^5 =
\gamma^0\gamma^1 = \sigma_3$, where $\sigma_i\,(i = 1,2,3)$ are Pauli
matrices.}. The products of fermion operators in the Lagrangian
(\ref{label1.1}) are taken in the normal--ordered form at the scale
$\mu \to 0$.

The massless Thirring model describes a system of self--coupled
fermions with a non--trivial four--fermion interaction. In quantum
field theoretic models, defined in 3+1--dimensional space--time
\cite{NJL}--\cite{AI01}, the four--fermion interaction is responsible
for a non--trivial vacuum with the properties of a BCS--type
superconducting ground state \cite{BCS}.

Recently we have shown \cite{FI1}--\cite{FI12} that the massless
Thirring model possesses a non--perturbative phase with the BCS--type
wave function \cite{FI1}.

It is well--known that the massless Thirring model is a solvable
model. All solutions of the massless Thirring model have been found in
the chiral symmetric phase \cite{KJ61}--\cite{KF}. The most general
solution of the massless Thirring model has been obtained by Hagen
\cite{CH67,FHI}. Hagen found a one--parameter family of solutions of
the massless Thirring model parameterized by the parameter $\xi$. For
$\xi = 1/2$ Hagen's solution coincides with Johnson's solution
\cite{KJ61}. For $\xi = 1$ Hagen's solution reproduces the results
obtained by Klaiber \cite{BK67}\,\footnote{Klaiber's regularization
procedure and parameterization has been recently critically discussed
in \cite{FI3}.} and within the path--integral approach \cite{KF}.

The massless Thirring model possesses four collective
fermion--antifermion excitations: $\psi^{\dagger}_1(x)\psi_1(x)$,
$\psi^{\dagger}_2(x)\psi_2(x)$, $\psi^{\dagger}_1(x)\psi_2(x)$ and
$\psi^{\dagger}_2(x)\psi_1(x)$. In a Lorentz covariant form they can
be written as
\begin{eqnarray}\label{label1.2}
A^0(x) &\sim& \bar{\psi}(x)\gamma^0\psi(x) =
\psi^{\dagger}_1(x)\psi_1(x) +
\psi^{\dagger}_2(x)\psi_2(x),\nonumber\\ A^1(x) &\sim&
\bar{\psi}(x)\gamma^1\psi(x) = \psi^{\dagger}_1(x)\psi_1(x) -
\psi^{\dagger}_2(x)\psi_2(x),\nonumber\\ \sigma(x) &\sim&
\bar{\psi}(x)\psi(x) = \psi^{\dagger}_2(x)\psi_1(x) +
\psi^{\dagger}_1(x)\psi_2(x),\nonumber\\ \varphi(x) &\sim&
\bar{\psi}(x)\gamma^5\psi(x) = \psi^{\dagger}_2(x)\psi_1(x) -
\psi^{\dagger}_1(x)\psi_2(x),
\end{eqnarray}
where $A^0(x)$ and $A^1(x)$ are the time and spatial components of the
2--vector $A^{\mu}(x) = (A^0(x), A^1(x))$; $\sigma(x)$ and
$\varphi(x)$ are chiral partners under chiral $U(1)\times U(1)$
rotations \cite{FI1}. In polar representation the collective
excitations $\sigma(x)$ and $\varphi(x)$ are defined by $\sigma(x) =
\rho(x)\,\cos\vartheta(x)$ and $\varphi(x) =
\rho(x)\,\sin\vartheta(x)$. In the chirally broken phase of the
massless Thirring model the vacuum expectation value of the field
$\rho(x)$ does not vanish and is equal to $\langle \rho(x) \rangle =
M$, where $M$ is the dynamical mass of Thirring fermion fields
quantized in the chirally broken phase \cite{FI1}\,\footnote{The
dynamical mass $M$ is equal to $M = \Lambda/
\sqrt{e^{\textstyle\,2\pi/g} - 1}$, where $\Lambda$ is the
ultra--violet cut--off \cite{FI1}.}. As has been shown in
\cite{FI1,FI12}, the evolution of fermions in the massless Thirring
model obeys the constant of motion
\begin{eqnarray}\label{label1.3}
:[\bar{\psi}(x)\psi(x)]^2 + [\bar{\psi}(x)i\gamma^5\psi(x)]^2: = C,
\end{eqnarray}
where $C = M^2/g^2$ \cite{FI1}.

In terms of the components of the fermion fields the four-fermion
interaction reads
\begin{eqnarray}\label{label1.4}
- \frac{1}{2}:\bar{\psi}(x)\gamma_{\mu}\psi(x)\bar{\psi}(x)
\gamma^{\mu}\psi(x): = -
2\,:\psi^{\dagger}_1(x)\psi_1(x)\psi^{\dagger}_2(x)\psi_2(x):.
\end{eqnarray}
The chiral invariant four--fermion interaction, containing the scalar
and pseudoscalar four--fermion vertices, is equal to
\begin{eqnarray}\label{label1.5}
\frac{1}{2}:(\bar{\psi}(x)\psi(x))^2 +
(\bar{\psi}(x)i\gamma^5\psi(x))^2: &=&
:\psi^{\dagger}_1(x)\psi_2(x)\psi^{\dagger}_2(x)\psi_1(x):\nonumber\\
&+& :\psi^{\dagger}_2(x)\psi_1(x)\psi^{\dagger}_1(x)\psi_2(x):.
\end{eqnarray}
Using equal--time canonical anti--commutation relations
\begin{eqnarray}\label{label1.6}
\{\psi_i(x^0,x^1),\psi^{\dagger}_j(x^0,y^1)\} = \delta_{ij}\delta(x^1
- y^1),
\end{eqnarray}
where $i(j) = 1,2$, we reduce the r.h.s. of (\ref{label1.5}) to the
form
\begin{eqnarray}\label{label1.7}
\frac{1}{2}:(\bar{\psi}(x)\psi(x))^2 +
(\bar{\psi}(x)i\gamma^5\psi(x))^2: &=& -\,2
:\psi^{\dagger}_1(x)\psi_1(x)\psi^{\dagger}_2(x)\psi_2(x):\nonumber\\
&+& \delta(0):\psi^{\dagger}_1(x)\psi_1(x) +
\psi^{\dagger}_2(x)\psi_2(x):
\end{eqnarray}
Using (\ref{label1.7}) the four--fermion interaction of Thirring
fermions can be transcribed into the form
\begin{eqnarray}\label{label1.8}
- \frac{1}{2}:\bar{\psi}(x)\gamma_{\mu}\psi(x)\bar{\psi}(x)
\gamma^{\mu}\psi(x): &=& \frac{1}{2}:(\bar{\psi}(x)\psi(x))^2 +
(\bar{\psi}(x)i\gamma^5\psi(x))^2:\nonumber\\ &-&
\delta(0):\bar{\psi}(x)\gamma^0\psi(x):.
\end{eqnarray}
Hence, one can conclude that the four--fermion interaction of Thirring
fermions is responsible for the description of vector, scalar and
pseudoscalar fermion--antifermion collective excitations.

Following Nambu and Jona--Lasino \cite{NJL} we introduce these
excitations with coupling constants $g_1$ and $g_2$.  However, in our
case, unlike the Nambu--Jona--Lasino approach \cite{NJL}, the coupling
constant $g_1$ and $g_2$ obey the constraint $g_1 + g_2 = g$. The
four--fermion interactions can then be rewritten in the following form
\begin{eqnarray}\label{label1.9}
&&-\frac{1}{2}\,g:\bar{\psi}(x)\gamma_{\mu}\psi(x)\bar{\psi}(x)
\gamma^{\mu}\psi(x): = -
\frac{1}{2}\,g_1:\bar{\psi}(x)\gamma_{\mu}\psi(x)\bar{\psi}(x)
\gamma^{\mu}\psi(x):\nonumber\\ &&+
\frac{1}{2}\,g_2:(\bar{\psi}(x)\psi(x))^2 + (\bar{\psi}(x)
i\gamma^5\psi(x))^2: - \,g_2\,\delta(0):\bar{\psi}(x)\gamma^0\psi(x):.
\end{eqnarray}
Due to (\ref{label1.9}) the Lagrangian (\ref{label1.1}) reads 
\begin{eqnarray}\label{label1.10}
{\cal L}_{\rm Th}(x) &=& :\bar{\psi}(x)(i\gamma^{\mu}\partial_{\mu} -
\,g_2\,\delta(0)\gamma^0)\psi(x): -
\frac{1}{2}\,g_1:\bar{\psi}(x)\gamma^{\mu}\psi(x)\bar{\psi}(x)
\gamma_{\mu}\psi(x):\nonumber\\ &+&
\frac{1}{2}\,g_2:(\bar{\psi}(x)\psi(x))^2 + (\bar{\psi}(x)
i\gamma^5\psi(x))^2:.
\end{eqnarray}
The term $g_2\,\delta(0)\gamma^0$ can be removed by a time--dependent
phase transformation $\psi(x) \to \psi\,'(x) = e^{\textstyle\,
-\,i g_2 \delta(0) x^0}\,\psi(x)$. This gives
\begin{eqnarray}\label{label1.11}
{\cal L}_{\rm Th}(x) &=&
:\bar{\psi}(x)i\gamma^{\mu}\partial_{\mu}\psi(x): -
\frac{1}{2}\,g_1:\bar{\psi}(x)\gamma^{\mu}\psi(x)\bar{\psi}(x)
\gamma_{\mu}\psi(x):\nonumber\\ &&+
\frac{1}{2}\,g_2:(\bar{\psi}(x)\psi(x))^2 + (\bar{\psi}(x)
i\gamma^5\psi(x))^2:.
\end{eqnarray}
The massless Thirring model can be represented in terms of local
fields of fermion--antifermion collective excitations within the
path--integral approach \cite{FI1}. In accordance with
(\ref{label1.11}), the generating functional of correlation
functions we define as \cite{FI1}
\begin{eqnarray}\label{label1.12}
Z_{\rm Th}[s,p,a_{\mu}] &=& \int {\cal D}\psi{\cal
D}\bar{\psi}\,\exp\,i\int d^2x\,\Big\{\bar{\psi}(x)
i\,\hat{\partial}\psi(x) -
\frac{1}{2}\,g_1\,\bar{\psi}(x)\gamma_{\mu}\psi(x)
\bar{\psi}(x)\gamma^{\mu}\psi(x)\nonumber\\&& +
\frac{1}{2}\,g_2\,(\bar{\psi}(x)\psi(x))^2 + \frac{1}{2}\,g_2\,
(\bar{\psi}(x)i\gamma^5\psi(x))^2 + \bar{\psi}(x)\psi(x)
s(x)\nonumber\\ && + \bar{\psi}(x)i\gamma^5\psi(x) p(x) +
\bar{\psi}(x)\gamma^{\mu}\psi(x)a_{\mu}(x)\Big\},
\end{eqnarray}
where $s(x)$, $p(x)$ and $a_{\mu}(x)$ are external sources of scalar,
pseudoscalar and vector fermion--antifermion densities. The generating
functional of correlation functions (\ref{label1.12}) is normalized by
$Z_{\rm Th}[0,0,0] = 1$.

In terms of $Z_{\rm Th}[s,p,a_{\mu}]$ the fermion condensate is
defined by
\begin{eqnarray}\label{label1.13}
\frac{1}{i}\,\frac{\delta Z_{\rm Th}[s,p,a_{\mu}]}{\delta
s(x)}\Bigg|_{s = p = a_{\mu} = 0} = \langle
\bar{\psi}(x)\psi(x)\rangle.
\end{eqnarray}
The linearization of the four--fermion interactions we carry out in
the usual way
\begin{eqnarray}\label{label1.14}
\hspace{-0.5in}&&Z_{\rm Th}[s,p,a_{\mu}] = \int {\cal D}\psi{\cal
D}\bar{\psi}\,\exp\,i\int d^2x\,\Big\{\bar{\psi}(x)(i\,\hat{\partial}
+ \hat{A}(x) + \hat{a}(x) - \sigma(x) + s(x)\nonumber\\
\hspace{-0.5in}&& - i\gamma^5\,\varphi(x) + i\gamma^5 p(x))\psi(x) +
\frac{1}{2g_1}\,A_{\mu}(x)A^{\mu}(x) - \frac{1}{2g_2}\,(\sigma^2(x) +
\varphi^2(x))\Big\}.
\end{eqnarray}
Making a change of variables $A_{\mu} + a_{\mu} \to A_{\mu}$, $\sigma
- s \to \sigma$ and $\varphi - p \to \varphi$ we get
\begin{eqnarray}\label{label1.15}
\hspace{-0.5in}&&Z_{\rm Th}[s,p,a_{\mu}] = \int {\cal D}\psi{\cal
D}\bar{\psi}\,\exp\,i\int d^2x\,\Big\{\bar{\psi}(x)(i\,\hat{\partial}
+ \hat{A}(x) - \sigma(x) - i\gamma^5\,\varphi(x))\psi(x)\nonumber\\
\hspace{-0.5in}&& + \frac{1}{2g_1}\,A_{\mu}(x)A^{\mu}(x) -
\frac{1}{2g_2}\,(\sigma^2(x) + \varphi^2(x)) -
\frac{1}{g_1}\,A_{\mu}(x)a^{\mu}(x) +
\frac{1}{2g_1}\,a_{\mu}(x)a^{\mu}(x) \nonumber\\
\hspace{-0.5in}&& - \frac{1}{g_2}\,\sigma(x)s(x) -
\frac{1}{g_2}\,\varphi(x)p(x) -
\frac{1}{2g_2}\,(s^2(x) + p^2(x))\Big\}.
\end{eqnarray}
Integrating over the fermion fields we arrive at the expression
\cite{FI1}
\begin{eqnarray}\label{label1.16}
\hspace{-0.3in}&&Z_{\rm Th}[s,p,a_{\mu}] = \int {\cal D}\sigma{\cal
D}\varphi{\cal D}^2A\exp\,i\int d^2x \Big\{{\cal L}_{\rm
eff}[\sigma(x),\varphi(x),A^{\mu}(x)] -
\frac{1}{g_1}\,A_{\mu}(x)a^{\mu}(x) \nonumber\\
\hspace{-0.3in}&&-
\frac{1}{g_2}\,\sigma(x)s(x) - \frac{1}{g_2}\,\varphi(x)p(x) +
\frac{1}{2g_1}\,a_{\mu}(x)a^{\mu}(x) - \frac{1}{2g_2}\,(s^2(x) +
p^2(x)) \Big\}.
\end{eqnarray}
The effective Lagrangian ${\cal L}_{\rm
eff}[\sigma(x),\varphi(x),A^{\mu}(x)]$ is defined by
\begin{eqnarray}\label{label1.17}
{\cal L}_{\rm eff}[\sigma(x),\varphi(x),A^{\mu}(x)] &=& -i\,{\rm
tr}\langle x|{\ell n}\,(i\,\hat{\partial} + \hat{A} - \sigma -
i\gamma^5\,\varphi)|x\rangle\nonumber\\ && +
\frac{1}{2g_1}\,A_{\mu}(x)A^{\mu}(x) - \frac{1}{2g_2}\,(\sigma^2(x) +
\varphi^2(x)),
\end{eqnarray}
where the first term is caused by the contribution of the fermion
determinant with the trace calculated over the Dirac matrices.

According to (\ref{label1.14}), we can write down the bosonization
rules \cite{FI1}
\begin{eqnarray}\label{label1.18}
\bar{\psi}(x)\gamma^{\mu}\psi(x) = -
\frac{A^{\mu}(x)}{g_1}\;,\;\bar{\psi}(x)\psi(x) = -
\frac{\sigma(x)}{g_2}\;,\;\bar{\psi}(x)i\gamma^5\psi(x) = -
\frac{\varphi(x)}{g_2}.
\end{eqnarray}
In terms of the generating functional (\ref{label1.16}) the fermion
condensate $\langle \bar{\psi}(x)\psi(x)\rangle$ is defined by
\begin{eqnarray}\label{label1.19}
\langle \bar{\psi}(x)\psi(x)\rangle = \frac{1}{i}\frac{\delta Z_{\rm
Th}[s,p,a_{\mu}]}{\delta s(x)}\Bigg|_{s = p = a_{\mu} = 0} =
-\frac{1}{g_2}\,\langle \sigma(x)\rangle.
\end{eqnarray}
This agrees with the bosonization rules (\ref{label1.18}).

The main question, which we would like to clarify below, is what kind
of collective excitations is the most important for the solution of
the massless Thirring model $A^{\mu}$ or $\sigma$ and $\varphi$? The
reply on this question can be obtained by investigating the effective
potential of the massless Thirring model in terms of the fields of the
collective excitations.

The paper is organized as follows. In Section 2 we calculate the
effective potential of the massless Thirring model in terms of the
local fields of fermion--antifermion collective excitations. We
investigate the minima of this potential and show that the absolute
minimum corresponds only to the existence of $\sigma$ and $\varphi$
collective excitations in agreement with our analysis of the massless
Thirring model carried out in \cite{FI1}--\cite{FI11}. In Section 3 we
discuss the wave function of the ground state of the massless Thirring
model around the absolute minimum of the effective potential of
fermion--antifermion collective excitations.  In the Conclusion we
discuss the obtained results.

\section{Effective potential of collective excitations in the 
massless Thirring model}
\setcounter{equation}{0}

It is well--known that the minima of the effective potential of
collective excitations define the ground states of the quantum system
-- the vacua. Therefore, for the understanding of the problem of the
vacua of the massless Thirring model we have to calculate the
effective potential as a functional of $\sigma(x)$, $\varphi(x)$ and
$A_{\mu}(x)$ fields, i.e. $V_{\rm eff}[\sigma,\varphi,A]$.

For the calculation of the effective potential $V_{\rm
eff}[\sigma,\varphi,A]$ of the quantum field theory defined by the
effective Lagrangian (\ref{label1.17}) we drop the contributions of
derivatives of the collective fields $\sigma$, $\varphi$ and
$A_{\mu}$. The general expression for the effective potential reads
\cite{FI1}
\begin{eqnarray}\label{label2.1}
V_{\rm eff}[\sigma,\varphi, A] = -\tilde{{\cal L}}_{\rm
eff}[\sigma,\varphi,A]|_{\partial \sigma = \partial
\varphi = \partial A = 0} - \frac{1}{2g_1}\,A_{\mu}(x)A^{\mu}(x) +
\frac{1}{2g_2}\,(\sigma^2(x) + \varphi^2(x)).
\end{eqnarray}
The effective Lagrangian $\tilde{{\cal L}}_{\rm eff}[\sigma,\varphi,
A]|_{\partial\sigma = \partial \varphi = \partial A = 0}$ is
determined by
\begin{eqnarray}\label{label2.2}
\hspace{-0.5in}\tilde{{\cal L}}_{\rm eff}[\sigma,\varphi,A]|_{\partial
\sigma = \partial \varphi = \partial A = 0} &=& \frac{1}{2i}\,{\rm
tr}\langle x|{\ell n}(-(i\,\partial + A)^2 +
\sigma^2+\varphi^2)|x\rangle =\nonumber\\ \hspace{-0.5in}&& =\int
\frac{d^2k}{(2\pi)^2i}\,{\ell n}(-(k + A(x))^2 +
\sigma^2+\varphi^2))=\nonumber\\ =\hspace{-0.5in}&&= \int
\frac{d^2k}{(2\pi)^2i}\,\exp\Big(\,A(x)\cdot \frac{\partial}{\partial
k}\Big)\,{\ell n}(- k^2 + \sigma^2+\varphi^2).
\end{eqnarray}
In the polar field variables $\sigma(x) = \rho(x)\,\cos\vartheta(x)$
and $\varphi(x) = \rho(x)\,\sin\vartheta(x)$ the effective potential
depends only on $\rho(x)$ and $A_{\mu}(x)$
\begin{eqnarray}\label{label2.3}
\hspace{-0.5in}&&V_{\rm eff}[\rho, A] = - \int
\frac{d^2k}{(2\pi)^2i}\,\exp\Big(\,A(x)\cdot \frac{\partial}{\partial
k}\Big)\,{\ell n}(- k^2 + \rho^2) - \frac{1}{2g_1}\,A^2(x) +
\frac{1}{2g_2}\,\rho^2(x),
\end{eqnarray}
where we have denoted $A^2(x) = A_{\mu}(x)A^{\mu}(x)$.

By a Wick rotation we obtain
\begin{eqnarray}\label{label2.4}
V_{\rm eff}[\rho,A_{\rm E}] &=& -\int \frac{d^2k_{\rm
E}}{(2\pi)^2}\,\exp\Big(\,A_{\rm E}(x)\cdot \frac{\partial}{\partial
k_{\rm E}}\Big)\,{\ell n}(k^2_{\rm E} + \rho^2(x)) +
\frac{1}{2g_1}\,A^2_{\rm E}(x) + \frac{1}{2g_2}\,\rho^2(x)
=\nonumber\\ &=&-\int \frac{d^2k_{\rm E}}{(2\pi)^2}\,{\ell n}((k_{\rm
E} + A_{\rm E})^2 + \rho^2(x)) + \frac{1}{2g_1}\,A^2_{\rm E}(x) +
\frac{1}{2g_2}\,\rho^2(x),
\end{eqnarray}
where $A^2_{\rm E}(x) = - A_{\mu}(x) A^{\mu}(x)$. 

The calculation of the integral over $k$ reads
\begin{eqnarray}\label{label2.5}
\hspace{-0.5in}&&\int \frac{d^2k}{(2\pi)^2}\, {\ell n}((k + A)^2 +
\rho^2) = \frac{1}{4\pi^2}\int^{\Lambda}_{0}dk\,k\int^{2\pi}_0
d\alpha\,{\ell n}(k^2 + \rho^2 + A^2 + 2kA\,\cos\alpha) =\nonumber\\
\hspace{-0.5in}&&=\frac{1}{8\pi^2}\int^{\Lambda^2}_{0}du\int^{2\pi}_0
d\alpha\,{\ell n}(u + \rho^2 + A^2 + 2\sqrt{u}\,A\,\cos\alpha) =
\nonumber\\
\hspace{-0.5in}&&=\frac{\Lambda^2}{8\pi^2}\,\int^{2\pi}_0
d\alpha\,{\ell n}(\Lambda^2 + \rho^2 + A^2 + 2 \Lambda
A\,\cos\alpha)\nonumber\\
\hspace{-0.5in}&& -
\frac{1}{8\pi^2}\int^{\Lambda^2}_{0}du\int^{2\pi}_0 d\alpha\,\frac{u
+ \sqrt{u}\,A\,\cos\alpha}{u + \rho^2 + A^2 +
2\sqrt{u}\,A\,\cos\alpha} =\nonumber\\
\hspace{-0.5in}&&=\frac{\Lambda^2}{8\pi^2}\,\int^{2\pi}_0
d\alpha\,{\ell n}(\Lambda^2 + \rho^2 + A^2 + 2 \Lambda A\,\cos\alpha)
- \frac{\Lambda^2}{8\pi}\nonumber\\
\hspace{-0.5in}&& - \frac{1}{16\pi^2}\int^{\Lambda^2}_{0}du\,(u -
\rho^2 - A^2)\int^{2\pi}_0 d\alpha\,\frac{1}{u + \rho^2 + A^2 +
2\sqrt{u}\,A\,\cos\alpha}.
\end{eqnarray}
In this equation $A = \sqrt{A^2_{\rm E}}$ and $\Lambda$ is the
ultra--violet cut--off.

In the last integral over $\alpha$ it is convenient
to make a change of variables $e^{\textstyle i\,\alpha} = z$ and
recast it into the contour integral
\begin{eqnarray}\label{label2.6}
&&\int^{2\pi}_0 d\alpha\,\frac{1}{u + \rho^2 + A^2 +
2\sqrt{u}a\,\cos\alpha}=\frac{2\pi}{A\,\sqrt{u}}\oint_{|z|=1}
\frac{dz}{2\pi i}\, \frac{1}{\displaystyle z^2 + z\,\frac{u + \rho^2 +
A^2}{A\,\sqrt{u}} + 1} = \nonumber\\
&&=\frac{2\pi}{a\sqrt{u}}\oint_{|z|=1}\frac{dz}{2\pi i}\, \frac{1}{(z
- z_1)\,(z - z_2)}= \frac{2\pi}{z_1 - z_2} = \frac{2\pi}{\sqrt{(u
+\rho^2 + A^2)^2 - 4 u A^2}} = \nonumber\\ &&=\frac{2\pi}{\sqrt{u^2 +
2(\rho^2 - A^2) u + (\rho^2 + A^2)^2}},
\end{eqnarray}
where $z_1$ and $z_2$ are two roots of the denominator
\begin{eqnarray}\label{label2.7}
z_1&=&-\frac{u + \rho^2+ A^2}{2A\sqrt{u}} + \frac{\sqrt{(u +\rho^2 +
A^2)^2 - 4 u A^2}}{2a\sqrt{u}},\nonumber\\ z_2&=&-\frac{u + \rho^2+
A^2}{2A\sqrt{u}} - \frac{\sqrt{(u +\rho^2 + A^2)^2 - 4 u
A^2}}{2A\sqrt{u}}
\end{eqnarray}
The effective potential is now defined by
\begin{eqnarray}\label{label2.8}
\hspace{-0.5in}&&V_{\rm eff}[\rho,A] =\frac{\Lambda^2}{8\pi}
-\frac{\Lambda^2}{8\pi^2}\,\int^{2\pi}_0 d\alpha\,{\ell n}(\Lambda^2 +
\rho^2 + A^2_{\rm E} + 2 \Lambda A_{\rm E}\,\cos\alpha)\nonumber\\
\hspace{-0.5in}&& +\frac{1}{8\pi}\int^{\Lambda^2}_0
du\,\frac{u - \rho^2 - A^2_{\rm E}}{\sqrt{u^2 + 2(\rho^2 - A^2_{\rm E}) u + (\rho^2 +
A^2_{\rm E})^2}} + \frac{1}{2g_1}\,A^2_{\rm E} +
\frac{1}{2g_2}\,\rho^2,
\end{eqnarray}
where $A_{\rm E} = \sqrt{-A_{\mu}A^{\mu}}$.

\begin{figure}[h]
\psfrag{xlab}{$\displaystyle{\rho/\Lambda}$}
\psfrag{ylabel}{$\hspace{-5mm}\displaystyle{\frac{4\pi}{\Lambda^2}} \;
V_{\rm eff}[\rho,0]$}
\centerline{\includegraphics[width=0.6\textwidth]{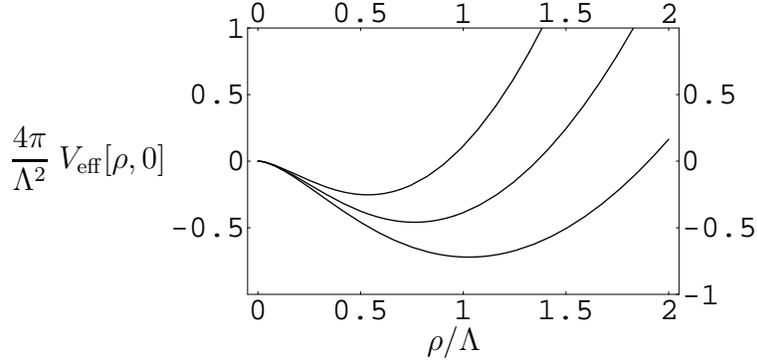}}
\caption{The effective potential $V_{\rm eff}[\rho,0]$ of
Eq.(\ref{label2.12}) as a function of $\rho/\Lambda$ for $2\pi/g = 2\pi/g_2 =
2/3$, $1$ and $3/2$. With increasing $g$ the minimum gets lower and is
shifted to larger values of $\rho$.}
\label{fig1}
\end{figure}

Integrating over $u$ we obtain
\begin{eqnarray}\label{label2.9}
&&\frac{1}{8\pi}\int^{\Lambda^2}_0 du\,\frac{u - \rho^2 - A^2_{\rm
E}}{\sqrt{u^2 + 2(\rho^2 - A^2_{\rm E}) u + (\rho^2 + A^2_{\rm E})^2}}
=\nonumber\\ &&= \frac{1}{8\pi}\int^{\Lambda^2}_0 du\,\frac{(u +\rho^2
- A^2_{\rm E}) - 2\rho^2}{\sqrt{(u^2 + \rho^2 - A^2_{\rm E})^2 +
4\rho^2 A^2_{\rm E}}}=\frac{1}{8\pi}\,\sqrt{(u^2 + \rho^2 - A^2_{\rm
E})^2 + 4\rho^2 A^2_{\rm E}}\Big|^{\Lambda^2}_0\nonumber\\ && -
\frac{\rho^2}{4\pi}\int^{\Lambda^2}_0 du\,\frac{1}{\sqrt{(u^2 + \rho^2
- A^2_{\rm E})^2 + 4\rho^2 A^2_{\rm E}}} =\nonumber\\
&&=\frac{1}{8\pi}\,\Big(\sqrt{(\Lambda^2 + \rho^2 - A^2_{\rm E})^2 +
4\rho^2 A^2_{\rm E}} - (\rho^2 + A^2_{\rm E})\Big)\nonumber\\ && -
\frac{\rho^2}{4\pi}\,{\ell n}\Bigg[\frac{\Lambda^2 + \rho^2 - A^2_{\rm
E} + \sqrt{(\Lambda^2 + \rho^2 - A^2_{\rm E})^2 + 4\rho^2 A^2_{\rm
E}}}{2\rho^2}\Bigg].
\end{eqnarray}
The integral over $\alpha$ can be calculated explicitly and the result
reads
\begin{eqnarray}\label{label2.10}
&&\int^{2\pi}_0 d\alpha\,{\ell n}(\Lambda^2 + \rho^2 + A^2_{\rm E} + 2
\Lambda A_{\rm E}\,\cos\alpha) = \nonumber\\ &&= 2\pi\,{\ell
n}\Bigg[\frac{\Lambda^2 + \rho^2 + A^2_{\rm E} + \sqrt{(\Lambda^2 +
\rho^2 + A^2_{\rm E})^2 - 4\Lambda^2 A^2_{\rm E}}}{2}\Bigg].
\end{eqnarray}
Substituting (\ref{label2.9}) and (\ref{label2.10}) in
(\ref{label2.8}) and imposing the condition $V_{\rm eff}[0,0] = 0$, we
end up with the expression for the effective potential 
\begin{eqnarray}\label{label2.11}
\hspace{-0.5in}V_{\rm eff}[\rho, A_{\rm E}]
&=&\frac{\Lambda^2}{4\pi}\,\Bigg\{ -{\ell n}\Bigg[\frac{1}{2}\Bigg(1 +
\frac{\rho^2}{\Lambda^2} + \frac{A^2_{\rm E}}{\Lambda^2} +
\sqrt{\Bigg(1 + \frac{\rho^2}{\Lambda^2} + \frac{A^2_{\rm
E}}{\Lambda^2}\Bigg)^2 - 4\,\frac{A^2_{\rm
E}}{\Lambda^2}}\;\Bigg)\Bigg]\nonumber\\
\hspace{-0.5in}&& - \frac{1}{2}\,\Bigg(1+ \frac{\rho^2}{\Lambda ^2} +
\frac{A^2_{\rm E}}{\Lambda ^2} - \sqrt{\Bigg(1 +
\frac{\rho^2}{\Lambda^2} - \frac{A^2_{\rm E}}{\Lambda^2}\Bigg)^2 +
4\,\frac{\rho^2}{\Lambda^2}\,\frac{A^2_{\rm E}}{\Lambda^2}}\Bigg) +
\frac{\rho^2}{\Lambda^2}\,{\ell n}\frac{\rho^2}{\Lambda^2}\nonumber\\
\hspace{-0.5in}&& - \frac{\rho^2}{\Lambda^2}\,{\ell
n}\Bigg[\frac{1}{2}\Bigg(1 + \frac{\rho^2}{\Lambda^2} - \frac{A^2_{\rm
E}}{\Lambda^2} + \sqrt{\Bigg(1 + \frac{\rho^2}{\Lambda^2} -
\frac{A^2_{\rm E}}{\Lambda^2}\Bigg)^2 +
4\,\frac{\rho^2}{\Lambda^2}\frac{A^2_{\rm
E}}{\Lambda^2}}\;\Bigg)\Bigg]\nonumber\\
\hspace{-0.5in}&& + \frac{2\pi}{g_1}\,\frac{A^2_{\rm E}}{\Lambda^2} +
\frac{2\pi}{g_2}\,\frac{\rho^2}{\Lambda^2}\Bigg\}.
\end{eqnarray}
In the limit $g_1 \to 0$ corresponding $A_{\rm E} \to 0$ we get the
effective potential $V_{\rm eff}[\rho,0]$
\begin{eqnarray}\label{label2.12}
\hspace{-0.5in}&&V_{\rm eff}[\rho,0]
=\frac{\Lambda^2}{4\pi}\Bigg[\frac{\rho^2}{\Lambda^2}\,{\ell
n}\frac{\rho^2}{\Lambda^2} - \Bigg(1 +
\frac{\rho^2}{\Lambda^2}\Bigg)\,{\ell n}\Bigg(1 +
\frac{\rho^2}{\Lambda^2}\Bigg) +
\frac{2\pi}{g}\,\frac{\rho^2}{\Lambda^2} \Bigg],
\end{eqnarray}
which coincides with the effective potential evaluated in \cite{FI1}
(see Eq.(3.20)). As we have already found in \cite{FI1}, the minimum
of the effective potential (\ref{label2.12}) is at $\rho = M$:
\begin{eqnarray}\label{label2.13}
V_{\rm eff}[M,0] = - \frac{\Lambda^2}{4\pi}\,{\ell n}\Bigg(1 +
\frac{M^2}{\Lambda^2}\Bigg) = \frac{\Lambda^2}{4\pi}\,{\ell n}\Bigg(1
- e^{\textstyle\,-2\pi/g}\Bigg).
\end{eqnarray}
The shape of the potential $V_{\rm eff}[\rho,0]$ is depicted in
Fig.~\ref{fig1} as a function of $\rho/\Lambda$ for $2\pi/g = 2/3, 1,
3/2$.

\begin{figure}[h]
\psfrag{xlab}{$\displaystyle{A_{\rm E}/\Lambda}$}
\psfrag{ylabel}{$\hspace{-5mm}\displaystyle{\frac{4\pi}{\Lambda^2}} \; V_{\rm eff}[0,A_{\rm E}]$}
\centerline{\includegraphics[width=0.6\textwidth]{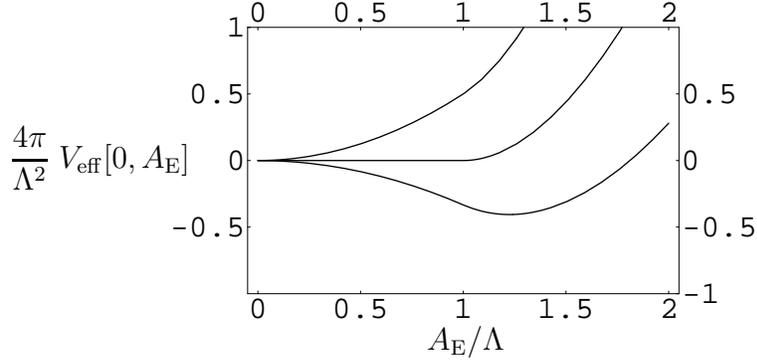}}
\caption{The effective potential $V_{\rm eff}[0,A_{\rm E}]$ of
Eq.(\ref{label2.12}) as a function of $A_{\rm E}/\Lambda$ for $2\pi/g =
2\pi/g_1 = 2/3$, $1$ and $3/2$. With increasing $g$ the potential is lowered.}
\label{fig2}
\end{figure}

In turn in the limit $g_2 \to 0$ leading to $\rho = 0$ we reduce the
effective potential (\ref{label2.11}) to the form
\begin{eqnarray}\label{label2.14}
\hspace{-0.3in}V_{\rm eff}[0, A_{\rm E}] &=&
\frac{\Lambda^2}{4\pi}\Bigg\{ -{\ell n}\Bigg[\frac{1}{2}\Bigg(1 +
\frac{A^2_{\rm E}}{\Lambda^2} + \Bigg|1 - \frac{A^2_{\rm
E}}{\Lambda^2}\Bigg|\Bigg)\Bigg]\nonumber\\
\hspace{-0.3in}&&- \frac{1}{2}\,\Bigg(1 + \frac{A^2_{\rm
E}}{\Lambda^2} - \Bigg|1 - \frac{A^2_{\rm E}}{\Lambda^2}\Bigg|\Bigg) +
\frac{2\pi}{g}\,\frac{A^2_{\rm E}}{\Lambda^2}\Bigg\}.
\end{eqnarray}
The shape of this potential can be seen in Fig.~\ref{fig2}.
Extrema of the effective potential (\ref{label2.14}) are defined
by the first derivative with respect to $A_{\rm E}$
\begin{eqnarray}\label{label2.15}
\hspace{-0.3in}\frac{\delta V_{\rm eff}[0, A_{\rm E}]}{\delta A_{\rm
E}} = \frac{1}{2\pi}\,\frac{A^2_{\rm E} - \Lambda^2}{A_{\rm
E}}\,\theta(A^2_{\rm E} - \Lambda^2) - \frac{1}{2\pi}\,\Big(1 -
\frac{2\pi}{g}\Big)\,A_{\rm E} = 0,
\end{eqnarray}
where $\theta(A^2_{\rm E} - \Lambda^2)$ is the Heaviside
function. The roots of the equation (\ref{label2.15}) are equal to
\begin{eqnarray}\label{label2.16}
A^{min}_{\rm E} = \left\{\begin{array}{r@{\;,\;}l} 0& A^2_{\rm E} <
\Lambda^2,\\ {\displaystyle \sqrt{\frac{g}{2\pi}}\,\Lambda} & A^2_{\rm
E} > \Lambda^2.
\end{array}\right.
\end{eqnarray}
A non--trivial solution of the equation (\ref{label2.15}) exists only
for $g > 2\pi$.

Analysing the second derivative of the effective potential
with respect to $A_{\rm E}$ we can obtain the conditions for the
existence of the minima. We get
\begin{eqnarray}\label{label2.17}
\hspace{-0.3in}\frac{\delta^2 V_{\rm eff}[0, A_{\rm E}]}{\delta
A^2_{\rm E}} = \frac{1}{2\pi}\,\frac{A^2_{\rm E} +\Lambda^2}{A^2_{\rm
E}}\,\theta(A^2_{\rm E} - \Lambda^2) - \frac{1}{2\pi}\,\Big(1 -
\frac{2\pi}{g}\Big).
\end{eqnarray}
We have used the relation $(A^2_{\rm E} - \Lambda^2)\delta(A^2_{\rm E}
- \Lambda^2) = 0$, where $\delta(A^2_{\rm E} - \Lambda^2)$ is the
Dirac $\delta$--function. For $A_{\rm E}$ given by (\ref{label2.16})
the second derivative is equal to
\begin{eqnarray}\label{label2.18}
\hspace{-0.3in}\frac{\delta^2 V_{\rm eff}[0, A^{min}_{\rm E}]}{\delta
A^2_{\rm E}} = \left\{\begin{array}{r@{\;,\;}l} {\displaystyle -
\frac{1}{2\pi} + \frac{1}{g}}& A^2_{\rm E} < \Lambda^2,\\
{\displaystyle \frac{2}{g}}~~~& A^2_{\rm E} > \Lambda^2.
\end{array}\right.
\end{eqnarray}
Hence, a minimum of the effective potential at $\rho = 0$ and $g_1 =
g$ can exist only either for coupling constants $g < 2\pi$ at
$A^{min}_{\rm E} = 0$ or for $g > 2\pi$ at $A^{min}_{\rm E} =
\sqrt{g/2\pi}\,\Lambda$.

We would like to emphasize that $A_{\rm E} = 0$ is not a real minimum
of the effective potential for $\rho = 0$ but a saddle--point. Indeed,
according to the criterion for the saddle--point, it is a minimum in
$A_{\rm E}$ but a maximum in $\rho$.

The effective potential (\ref{label2.14}) at $A^{min}_{\rm E}$, given by
(\ref{label2.16}), is equal to
\begin{eqnarray}\label{label2.19}
V_{\rm eff}[0, A^{min}_{\rm E}] = \left\{\begin{array}{r@{\;,\;}l}
0\hspace{0.5in}& A^{min}_{\rm E} = 0,\\ {\displaystyle
-\frac{\Lambda^2}{4\pi}\,{\ell n}\Big(\frac{g}{2\pi}\Big)} &
{\displaystyle A^{min}_{\rm E} = \sqrt{\frac{g}{2\pi}}\,\Lambda}.
\end{array}\right.
\end{eqnarray}
This could testify that the chiral symmetric phase of the massless
Thirring model can exist for coupling constants $g < 2\pi$ only.  But
$\rho = A_{\rm E} = 0$ is a saddle-point of the effective potential
for any $0 < g < 2\pi$. This is clearly seen from
Eqs.~(\ref{label2.12})--(\ref{label2.19}) and also Figs.~\ref{fig1}
and \ref{fig2}. Since the point $\rho = A_{\rm E} = 0$ is not a
minimum of the effective potential (\ref{label2.11}), the chiral
symmetric phase is not stable, and it should make a transition to the
chirally broken phase in the region $0 < g < 2\pi$. This agrees with
our results obtained in \cite{FI1,FI3}.

\begin{figure}[h]
\begin{picture}(0,0)
\put(250,100){$V_{\rm eff}[0,A^{\rm min}_{\rm E}]$}
\put(175,80){$V_{\rm eff}[M,0]$}
\end{picture}
\psfrag{g}{$\displaystyle{g}$}
\psfrag{p}{$\displaystyle{\pi}$}
\psfrag{Vmin}{$\displaystyle{V_{min}}$}
\centerline{\includegraphics[width=0.6\textwidth]{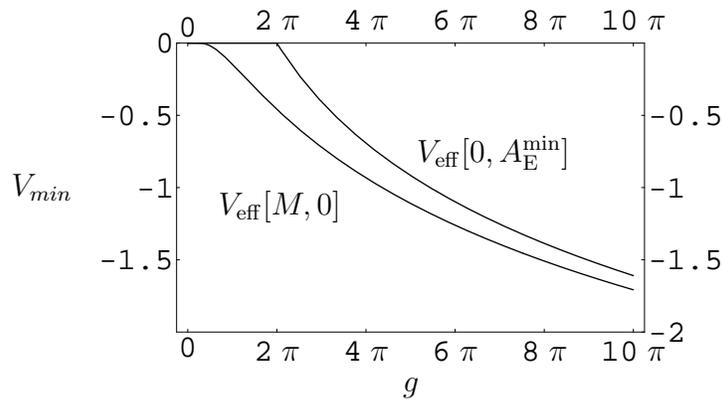}}
\caption{Depth of the minima $V_{\rm eff}[M,0]$ and $V_{\rm
eff}[0,A^{\rm min}_{\rm E}]$ of the effective potentials $V_{\rm
eff}[\rho,0]$ and $V_{\rm eff}[0,A_{\rm E}]$. The minimum of $V_{\rm
eff}[0,A_{\rm E}]$ is non-trivial for $g > 2\pi$ only. For all
couplings $g > 0$ the absolute minimum is the chirally broken minimum
$V_{\rm eff}[M,0]$.}
\label{fig3}
\end{figure}

The difference $\Delta V_{\rm eff}$ of the effective potentials
$V_{\rm eff}[M,0]$ and $V_{\rm eff}[0,A^{min}_{\rm E}]$ is equal to
\begin{eqnarray}\label{label2.20}
\Delta V_{\rm eff} = V_{\rm eff}[M,0] - V_{\rm eff}[0, A^{min}_{\rm
E}] = \left\{\begin{array}{r@{\;,\;}l}{\displaystyle
\frac{\Lambda^2}{4\pi}\,{\ell n}\Big(1 -
e^{\textstyle\,-2\pi/g}\,\Big)}& 0 < g < 2\pi,\\ {\displaystyle
\frac{\Lambda^2}{4\pi}\,{\ell n}\Big[\frac{g}{2\pi}\Big(1 -
e^{\textstyle\,-2\pi/g}\,\Big)\Big]} & g > 2\pi.
\end{array}\right.
\end{eqnarray}
It is seen that for $g > 0$ the minimum of the effective potential
$V_{\rm eff}[\rho,0]$ is always deeper than the minimum of $V_{\rm
eff}[0,A_{\rm E}]$. Therefore, below we call a metastable minimum and
an absolute minimum the values of the effective potential
(\ref{label2.11}) at $(\rho = 0, A_{\rm E} = \sqrt{g/2\pi}\Lambda)$
and $(\rho = M, A_{\rm E} = 0)$, respectively. These minima are
separated by a saddle-point in the $(\rho,A_{\rm E},\arctan(g_1/g_2)
)$ space for $g > 2\pi$. The values of these minima are depicted in
Fig.~\ref{fig3}.

\begin{figure}[h]
\begin{picture}(0,0)
\put(330,50){$\rho/\Lambda$}
\put(130,40){$A_{\rm E}/\Lambda$}
\put(45,180){$\arctan(g_1/g_2)$}
\end{picture}
\centerline{\includegraphics[width=0.6\textwidth]{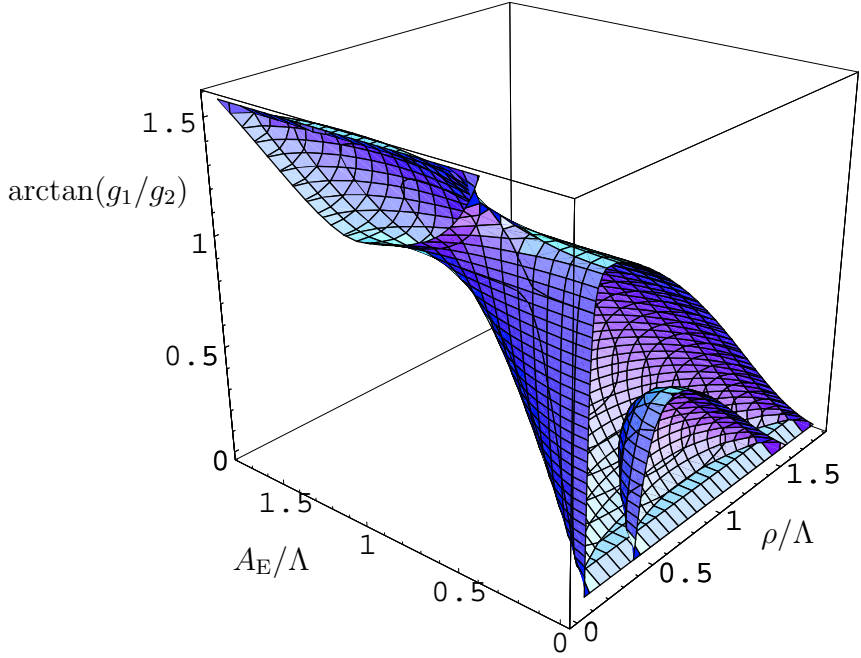}}
\caption{Effective potential for $g=3\pi$, depicted as a function of
$\rho/\Lambda$, $A_{\rm E}/\Lambda$ and $\arctan(g_1/g_2)$ by two of
the most characteristic equipotential surfaces above the absolute
minimum $V_{\rm eff}[M,0]$. For the lower value we have chosen $V_{\rm
eff}[\rho,A_{\rm E}] = V_{\rm eff}[M,0] + \Lambda^2/12\pi$ which
corresponds to the smaller surface.  The higher value $V_{\rm
eff}[M,0] + \Lambda^2/6\pi$ leads to an equipotential surface
connecting the two minima by a tunnel.}
\label{fig4}
\end{figure}

In Fig.~\ref{fig4} we give a graphical representation of the effective
potential (\ref{label2.11}) for $g=3\pi$ as a function of
$\rho/\Lambda$, $A_{\rm E}/\Lambda$ and $\arctan(g_1/g_2)$. The
variable $\arctan(g_1/g_2)$ is more convenient for the graphical
description of the system of fermion--antifermion collective
excitations. We have $\arctan(g_1/g_2) = 0$ for $g_1 = 0$ and $g_2 =
g$ (scalar and pseudoscalar degrees of freedom) and $\arctan(g_1/g_2)
= \pi/2$ for $g_1 = g$ and $g_2 = 0$ (vector degrees of freedom).

We illustrate the behaviour of the effective potential
(\ref{label2.11}) by two equipotential surfaces $V_{\rm eff}[\rho,
A_{\rm E}]={\rm const}$ above the absolute minimum at $(\rho,A_{\rm
E},\arctan(g_1/g_2))=(M,0,0)$ with $M/\Lambda = 1.03$. The
equipotential surface at $V_{\rm eff}[\rho,A_{\rm E}] = V_{\rm
eff}[M,0] + \Lambda^2/12\pi$ is below the value of the metastable
minimum at $(0,\sqrt{g/2\pi}\,\Lambda,\pi/2)$ and surrounds the
absolute minimum only.  The surface at $V_{\rm eff}[M,0] +
\Lambda^2/6\pi$ encloses both minima and connects them by a tunnel. At
the most narrow point of the tunnel there is a saddle--point of the
effective potential.  For the decreasing values of $g$ the
saddle--point moves towards $(0,0,\pi/2)$.

The non--zero value $A_{\rm E} = \sqrt{g/2\pi}\,\Lambda$ for $g >
2\pi$ should correspond to a chirally broken phase. This can be proved
by calculating the vacuum expectation value of the product of the
vector currents
\begin{eqnarray}\label{label2.21}
A^2_{\rm E} &=& - \langle A_{\mu}(x)A^{\mu}(x)\rangle = - g^2 \langle
\bar{\psi}(x)\gamma_{\mu}\psi(x)\bar{\psi}(x)\gamma_{\mu}\psi(x)\rangle
=\nonumber\\ &=& - g^2{\rm
tr}\{S_F(0)\gamma_{\mu}S_F(0)\gamma^{\mu}\}.
\end{eqnarray}
where we have used the bosonization rules (\ref{label1.18}) and
$S_F(x)$ is the two--point causal Green function of the Thirring
fermion fields, which we define using the K\"allen--Lehmann
representation \cite{FI11,FI12}
\begin{eqnarray}\label{label2.22}
\hspace{-0.3in}S_F(x)_{ab} = i\langle {\rm
T}(\psi_a(x)\bar{\psi}_b(0))\rangle = \int
\frac{d^2k}{(2\pi)^2}\,e^{\textstyle\,-ik\cdot
x}\int^{\infty}_0dm^2\,\frac{(\hat{k}\rho_1(m^2) +
\rho_2(m^2))_{ab}}{m^2 - k^2 - i\,0}.
\end{eqnarray}
Here $\rho_1(m^2)$ and $\rho_2(m^2)$ are K\"allen--Lehmann spectral
functions. In the chiral symmetric phase the spectral function
$\rho_2(m^2)$ should vanish, $\rho_2(m^2) = 0$. Let us show that the
non--zero value of $A_{\rm E}$ is defined only by $\rho_2(m^2) \neq
0$.

Substituting (\ref{label2.22}) in (\ref{label2.21}), using the identity
$\gamma_{\mu}\gamma^{\alpha}\gamma^{\mu} = 0$ and integrating over $k$
we end up with the expression \cite{FI11}
\begin{eqnarray}\label{label2.23}
A^2_{\rm E} =\Bigg[\frac{g}{2\pi}\int^{\infty}_0dm^2\,\rho_2(m^2)\,{\ell
n}\Big(1 + \frac{\Lambda^2}{m^2}\Big)\Bigg]^2.
\end{eqnarray}
Hence, $A_{\rm E} \neq 0$ only for $\rho_2(m^2) \neq 0$.  

For the comparison with $A_{\rm E}$ we suggest to calculate $\rho$
defined by
\begin{eqnarray}\label{label2.24}
\rho^2 &=& \langle \sigma^2(x) + \varphi^2(x)\rangle = g^2\langle
([\bar{\psi}(x)\psi(x)]^2 + [\bar{\psi}(x)i\gamma^5\psi(x)]^2)\rangle
= \nonumber\\ &=&g^2[{\rm tr}\{S_F(0)\}]^2 + g^2{\rm tr}\{S_F(0)S_F(0)
- \gamma^5S_F(0)\gamma^5S_F(0)\} = \nonumber\\
&=&\Bigg[\frac{g}{2\pi}\int^{\infty}_0dm^2\,\rho_2(m^2)\,{\ell
n}\Big(1 + \frac{\Lambda^2}{m^2}\Big)\Bigg]^2,
\end{eqnarray}
where we have also used the bosonization rules (\ref{label1.18}).
This testifies that only the saddle--point of the effective potential
$(\rho = 0, A_{\rm E} = 0)$ corresponds to the chiral symmetric phase
of the massless Thirring model. It can be realized only for
$\rho_2(m^2) = 0$. In turn, the minima $(\rho = 0, A_{\rm E} =
\sqrt{g/2\pi}\Lambda)$ and $(\rho = M, A_{\rm E} = 0)$ correspond to
the chirally broken phases of the massless Thirring model, which
demands $\rho_2(m^2) \neq 0$. The stability  of the minimum $(\rho = M,
A_{\rm E} = 0)$ under quantum fluctuations $\rho(x) = M +
\tilde{\rho}(x)$, where $\tilde{\rho}(x)$ is a fluctuating field has
been shown in \cite{FI1}.

\section{Ground state of the massless Thirring model at the minimum 
$(\rho = M, A_{\rm E} = 0)$}
\setcounter{equation}{0}

Placing massless Thirring fermions into a finite volume $L$ the BCS
wave function of the ground state of the massless Thirring model is
defined by \cite{FI1,FI6}
\begin{eqnarray}\label{label3.1}
|\Omega(0)\rangle = \prod_{p^1}[u_{p^1} +
 v_{p^1}\,a^{\dagger}(p^1)b^{\dagger}(-p^1)]\,|\Psi_0\rangle,
\end{eqnarray}
where $|\Psi_0\rangle$ is the wave function of the chiral symmetric
vacuum, the coefficients $u_{p^1}$ and $v_{p^1}$ have the properties:
(i) $u^2_{p^1} + v^2_{p^1} = 1$ and (ii) $u_{- p^1} = u_{p^1}$ and
$v_{-p^1} = - v_{p^1}$ \cite{FI1,FI6}, $a^{\dagger}(p^1)$ and
$b^{\dagger}(p^1)$ are creation operators of fermions and antifermions
with momentum $p^1$.

According to Nambu and Jona--Lasinio \cite{YN60}, the wave function
(\ref{label3.1}) should be a linear superposition of the
eigenfunctions $|\Omega_{2n}\rangle$ of the chirality operator $X$,
defined by
\begin{eqnarray}\label{label3.2}
X(x^0) = \lim_{L\to \infty}\int^{+ L/2}_{-
L/2}dx^1\,\psi^{\dagger}(x^0,x^1) \gamma^5 \psi(x^0,x^1),
\end{eqnarray}
with eigenvalues $X_n = 2n\,, n\in {\mathbb{Z}}$, i.e.
\begin{eqnarray}\label{label3.3}
|\Omega(0)\rangle =\sum_{n \in
 \mathbb{Z}}C_{2n}|\Omega_{2n}\rangle = \sum_{n \in \mathbb{Z}}
 |\Phi_{2n}\rangle.
\end{eqnarray}
A chiral rotation of a fermion field $\psi(x)$ is defined by
\cite{YN60} (see also \cite{FI1})
\begin{eqnarray}\label{label3.4}
e^{\textstyle\,-i\alpha_AX(x^0)}\,\psi(x)\,
e^{\textstyle\,+i\alpha_AX(x^0)} =
e^{\textstyle\,+i\gamma^5\alpha_A}\psi(x),
\end{eqnarray}
where we have used canonical anticommutation relations
\begin{eqnarray}\label{label3.5}
\{\psi(x^0,x^1),\psi^{\dagger}(x^0,y^1)\} = \delta(x^1 - y^1).
\end{eqnarray}
For chiral rotations of fermion fields with a chiral phase
$\alpha_{\rm A}$ the wave function (\ref{label3.1}) changes as follows
\cite{FI1}
\begin{eqnarray}\label{label3.6}
|\Omega(\alpha_{A})\rangle = \prod_{p^1}[u_{p^1} +
 v_{p^1}\,e^{\textstyle -2i\varepsilon(p^1)\alpha_{A}}\,
 a^{\dagger}(p^1)b^{\dagger}(-p^1)]\,|\Psi_0\rangle,
\end{eqnarray}
where $\varepsilon(p^1)$ is a sign function. In terms of
$|\Omega(\alpha_{A})\rangle$ the wave functions $|\Phi_{2n}\rangle$
are defined by \cite{YN60}
\begin{eqnarray}\label{label3.7}
|\Phi_{2n}\rangle = C_{2n}|\Omega_{2n}\rangle =
 \int^{2\pi}_0\frac{d\alpha_{A}}{2\pi}\,e^{\textstyle +2in\alpha_{\rm
 A}}\,|\Omega(\alpha_{A})\rangle.
\end{eqnarray}
Substituting (\ref{label3.7}) in (\ref{label3.6}) and using the
identity \cite{IG64}
\begin{eqnarray}\label{label3.8}
\sum_{n\in \mathbb{Z}}e^{\textstyle\,2in\alpha_{\rm A}} =
\pi\sum_{k\in \mathbb{Z}}\delta(\alpha_{\rm A} - k\pi)
\end{eqnarray}
one arrives at the BCS wave function (\ref{label3.1}). Defining the
$\delta$--functions as \cite{IG64}
\begin{eqnarray}\label{label3.9}
\delta(\alpha_{\rm A} - k\pi) = \lim_{\varepsilon \to
0}\,\frac{1}{\pi}\,\frac{\varepsilon}{(\alpha_{\rm A} - k\pi)^2 +
\varepsilon^2}
\end{eqnarray}
we obtain
\begin{eqnarray}\label{label3.10}
&&\sum^{\infty}_{n = - \infty} |\Phi_{2n}\rangle =
\int^{2\pi}_0\frac{d\alpha_{A}}{2\pi}\,\sum^{\infty}_{n = -
\infty}e^{\textstyle +2in\alpha_{\rm A}}\,|\Omega(\alpha_{A})\rangle =
\nonumber\\ &&= \frac{1}{2}\int^{2\pi}_0 d\alpha_{A}\,\sum^{\infty}_{k
= -\infty}\delta(\alpha_A - k\pi)|\Omega(\alpha_{A})\rangle =
\frac{1}{2}\int^{2\pi}_0 d\alpha_{A}\,\delta(\alpha_A)
|\Omega(\alpha_{A})\rangle\nonumber\\ &&+ \frac{1}{2}\int^{2\pi}_0
d\alpha_{A}\, \delta(\alpha_A - \pi)|\Omega(\alpha_{A})\rangle +
\frac{1}{2}\int^{2\pi}_0 d\alpha_{A}\, \delta(\alpha_A -
2\pi)|\Omega(\alpha_{A})\rangle =\nonumber\\ &&=
\frac{1}{4}\,|\Omega(0)\rangle + \frac{1}{2}\,|\Omega(\pi)\rangle +
\frac{1}{4}\,|\Omega(2\pi)\rangle = |\Omega(0)\rangle.
\end{eqnarray}
Since $|\Phi_{2n}\rangle$ are eigenfunctions of the chirality
operator, $X|\Phi_{2n}\rangle = 2n|\Phi_{2n}\rangle$, they transform
under chiral rotations as follows
\begin{eqnarray}\label{label3.11}
e^{\textstyle\,-i\alpha X(0)}|\Phi_{2n}\rangle &=&
 \int^{2\pi}_0\frac{d\alpha_{A}}{2\pi}\,e^{\textstyle +2in\alpha_{\rm
 A}}\,e^{\textstyle\,-i\alpha X(0)}|\Omega(\alpha_{A})\rangle =
 \nonumber\\ &=& \int^{2\pi}_0\frac{d\alpha_{A}}{2\pi}\,e^{\textstyle
 +2in\alpha_{\rm A}}\,|\Omega(\alpha_{A} + \alpha )\rangle =
 e^{\textstyle\,- i2n\alpha}|\Phi_{2n}\rangle.
\end{eqnarray}
This agrees with Nambu and Jona--Lasinio \cite{YN60}. 

It is seen that only the wave function $|\Phi_0\rangle$ with chirality
zero is invariant under chiral rotations. Hence, it can describe a
non--trivial chiral symmetric ground state of the massless Thirring
model in the vicinity of the minimum $(\rho = M, A_{\rm E} = 0)$.

As has been shown in \cite{FI1} the effective potential, calculated at
the minimum $(\rho = M, A_{\rm E} = 0)$, coincides with the energy
density of the massless Thirring model with the ground state described
by the wave function (\ref{label3.1}). The Hamiltonian of the massless
Thirring model is equal to \cite{FI1}
\begin{eqnarray}\label{label3.12}
{\cal H}(x^0,x^1) &=& -
:\bar{\psi}(x^0,x^1)i\gamma^1\frac{\partial}{\partial x^1}\psi(x^0,
x^1):_{\mu}\nonumber\\ && + \frac{1}{2}\,g:\bar{\psi}(x^0,x^1)
\gamma_{\mu} \psi(x^0,x^1) \bar{\psi}(x^0,x^1) \gamma^{\mu}
\psi(x^0,x^1):_{\mu},
\end{eqnarray}
where $:\ldots:$ indicates normal ordering at an infrared scale
$\mu$, which should be finally set to zero $\mu \to 0$ \cite{FI6}.

The energy density of massless Thirring fermions in the ground
state, described by the wave function (\ref{label3.1}), is defined by
\cite{FI1}
\begin{eqnarray}\label{label3.13}
{\cal E}(M) = \lim_{L\to \infty} \frac{1}{L}\int^{+L/2}_{-L/2}dx^1\,
\langle\Omega(0)|{\cal H}(x^0,x^1)|\Omega(0)\rangle.
\end{eqnarray}
As has been shown in \cite{FI1} it is equal to
\begin{eqnarray}\label{label3.14}
{\cal E}(M) = V[M,0],
\end{eqnarray}
where $V[M,0]$ is the effective potential of collective
fermion--antifermion excitations calculated at $(\rho = M, A_{\rm E} =
0)$.

Since the Hamiltonian (\ref{label3.12}) is invariant under chiral
rotations, the same energy density (\ref{label3.14}) corresponds to
the ground state described by the wave function
$|\Omega(\alpha_A)\rangle$ (\ref{label3.6})
\begin{eqnarray}\label{label3.15}
{\cal E}(M)_{\alpha_A} &=& \lim_{L\to \infty}
\frac{1}{L}\int^{+L/2}_{-L/2}dx^1\,
\langle\Omega(\alpha_A)|{\cal
H}(x^0,x^1)|\Omega(\alpha_A)\rangle = \nonumber\\ &=&\lim_{L\to
\infty} \frac{1}{L}\int^{+L/2}_{-L/2}dx^1\,
\langle\Omega(0)|e^{\textstyle\,+i\alpha_AX}\,{\cal
H}(x^0,x^1)\,e^{\textstyle\,-i\alpha_AX}\,|\Omega(0)\rangle
=\nonumber\\ &=& \lim_{L\to \infty}
\frac{1}{L}\int^{+L/2}_{-L/2}dx^1\, \langle\Omega(0)|{\cal
H}(x^0,x^1)|\Omega(0)\rangle = {\cal E}(M),
\end{eqnarray}
where we have used that
\begin{eqnarray}\label{label3.16}
e^{\textstyle\,+i\alpha_AX}\,{\cal
H}(x^0,x^1)\,e^{\textstyle\,-i\alpha_AX} = {\cal H}(x^0,x^1).
\end{eqnarray}
The fact that the BCS wave function is not an eigenstate of chirality
guarantees that the properties of the ground state do not depend on
the exact number of fermion pairs. 

In order to clarify this assertion we would like to draw a similarity
between chirality in the massless Thirring model with triality in QCD
\cite{MF95}. In QCD there exist no triality changing transitions, this
means a dynamical change of triality is impossible. It is well--known
that the confined phase in QCD is $Z(3)$ symmetric. Triality zero
states are screened, and triality non--zero states are confined. This
means that states with different triality behave differently. Whereas
in the high--temperature phase of QCD all triality states behave in
the same way, they get screened. This is guaranteed by the spontaneous
breaking of $Z(3)$ symmetry. In our case the situation is similar to
the deconfined phase. In the massless Thirring model there are no
chirality changing transitions. The ground state is of BCS-type,
defining a condensate of fermion-antifermion pairs with different
chiralities. In order to get a ground state with properties
independent on the exact value of total chirality of all
fermion--antifermion pairs, we need spontaneous breaking of chiral
symmetry similar to the the spontaneous breaking of $Z(3)$ symmetry in
QCD. Such a spontaneous breaking of chiral symmetry is realized by the
BCS wave function.

However, for the description of the ground state of massless Thirring
fermions around the minimum of the effective potential $(\rho = M,
A_{\rm E} = 0)$ one can use $|\Phi_{2n}\rangle$, the eigenfunction of
chirality operator $X$. One can show that the energy density of
massless Thirring fermions in the ground state, described by the wave
function $|\Phi_{2n}\rangle$, is equal to ${\cal E}(M) = V[M,0]$. The
proof runs in the way
\begin{eqnarray}\label{label3.17}
\hspace{-0.3in}&&{\cal E}(M)_{2n} = \lim_{L\to \infty}
\frac{1}{L}\int^{+L/2}_{-L/2}dx^1\, \langle \Phi_{2n}|{\cal
H}(x^0,x^1)|\Phi_{2n}\rangle = \lim_{L\to
\infty}\int^{2\pi}_0\frac{d\alpha\,'_A}{2\pi}\int^{2\pi}_0
\frac{d\alpha_A}{2\pi}\,\nonumber\\ \hspace{-0.3in}&&\times\,
e^{\textstyle\,-i 2 n (\alpha\,'_A -
\alpha_A)}\,\frac{1}{L}\int^{+L/2}_{-L/2}dx^1\,
\langle\Omega(\alpha\,'_A )|{\cal H}(x^0,x^1)|\Omega(\alpha_A)\rangle.
\end{eqnarray}
The wave functions $|\Omega(\alpha\,'_A)\rangle$ and
$|\Omega(\alpha_A)\rangle$ are orthogonal for $\alpha\,'_{\rm A} \neq
\alpha_{\rm A}$\cite{FI1}. Indeed, the scalar product $\langle
\Omega(\alpha\,'_{\rm A})|\Omega(\alpha_{\rm A})\rangle$ of the wave
function for $\alpha\,'_{\rm A}\neq \alpha_{\rm A}$ is equal to
\cite{FI1,YN60}
\begin{eqnarray}\label{label3.18}
\langle \Omega(\alpha\,'_A)|\Omega(\alpha_A)\rangle &=& \lim_{L \to
\infty} \exp\Bigg\{\frac{L}{2\pi}\int^{\infty}_0dk^1\,{\ell n}\Bigg[1
- \sin^2(\alpha\,'_{\rm A} - \alpha_{\rm A})\frac{M^2}{M^2 +
(k^1)^2}\Bigg]\Bigg\} =\nonumber\\ &=& \delta_{\alpha\,'_A \alpha_A}.
\end{eqnarray}
It is obvious that the matrix element of the Hamiltonian should be
diagonal
\begin{eqnarray}\label{label3.19}
\langle\Omega(\alpha\,'_A )|{\cal
H}(x^0,x^1)|\Omega(\alpha_A)\rangle = {\cal E}(M)\,
\delta_{\alpha\,'_A \alpha_A}.
\end{eqnarray}
This agrees with Nambu and Jona--Lasinio \cite{YN60}. Insertion of
(\ref{label3.19}) into (\ref{label3.17}) gives
\begin{eqnarray}\label{label3.20}
{\cal E}(M)_{2n} = {\cal E}(M).
\end{eqnarray}
The order parameter of the superconducting BCS ground state is defined
by the vacuum expectation value \cite{FI1}
\begin{eqnarray}\label{label3.21}
\langle {\cal O}_+\rangle &=& \lim_{L\to \infty}\frac{2}{L}
\sum_{p^1}\varepsilon(p^1)\langle
\Omega(0)|b^{\dagger}(-p^1)a^{\dagger}(p^1)|\Omega(0) \rangle
=\nonumber\\ &=&- \lim_{L \to \infty}\frac{2}{L}
\sum_{p^1}\varepsilon(p^1) u_{p^1} v_{p^1} = - \frac{M}{g}.
\end{eqnarray}
Using the procedure developed in \cite{FI8} one can show that in the
limit $L \to \infty$ the operator ${\cal O}_+$ coincides with the
scalar fermion density operator $\bar{\psi}(0)\psi(0)$.

It is important to emphasize that the operator ${\cal O}_+$ satisfies
the selection rules $\Delta X = 2$. This means that a non--vanishing
value of the order parameter can be obtained only for the ground state
of the massless Thirring fermions described by the BCS--type wave
functions (\ref{label3.1}) and (\ref{label3.6}), which are not
eigenfunctions of the chirality operator (\ref{label3.2}).

We would like to notice that spontaneous breaking of chiral symmetry
in the massless Thirring model is not due to the fermion condensation
but caused by the wave functions of the ground state of the fermion
system. Indeed, the vacuum expectation value of the operator ${\cal
O}_+$ is zero for any wave function $|\Phi_{2n}\rangle$ accepted as
the wave function of the ground state of the massless Thirring
model. The main peculiarity of the wave function $|\Phi_{2n}\rangle$
is the fixed chirality corresponding the fixed number of
fermion--antifermion pairs. As we have shown above the wave functions
of fermion system with a fixed number of fermion--antifermion pairs
describe the ground state with a broken chiral symmetry and Thirring
fermions with a non--vanishing dynamical mass $M$ but the fermion
condensate is zero.

However, the ground state (or the vacuum state) of quantum field
theories of fermions in 2D--dimensional space--time does not have a
fixed number of fermion--antifermion pairs but the number of them is
infinite. Therefore, the correct wave function of the vacuum (the
ground state) should be a linear superposition over all states with a
fixed chirality or the fixed number of fermion--antifermion
pairs. This leads to the BCS--type wave function. For the BCS--type
wave function the fermion condensate does not vanish
\cite{FI1}.

\section{Conclusion}
\setcounter{equation}{0}

We have analysed the massless Thirring model in terms
of local fields of collective fermion--antifermion excitations, which
can be excited by the four--fermion interaction. We have calculated
the effective potential $V_{\rm eff}[\rho,A_{\rm E}]$, where we have
denoted $A_{\rm E} = \sqrt{ - A_{\mu}(x)A^{\mu}(x)}$ and $\rho =
\sqrt{\sigma^2(x) + \varphi^2(x)}$. The fields $A^{\mu}(x)$,
$\sigma(x)$ and $\varphi(x)$ are the local fields of
fermion--antifermion collective excitations, related to Thirring
fermion fields through the bosonization rules given by
(\ref{label1.18}).

Setting $\rho = 0$ and $g_2 = 0$, we have shown that for coupling
constants $g < 2\pi$ the effective potential $V_{\rm eff}[0,A_{\rm
  E}]$ possesses a chiral symmetric minimum at $A_{\rm E} = 0$.
Allowing for $\rho \ne 0$ this state turns out to be only a
saddle--point. Hence, this state cannot define the real ground state
of the massless Thirring model which can be found at $(\rho = M,
A_{\rm E} = 0)$.

For $g > 2\pi$ there are two minima $(\rho = M, A_{\rm E} = 0)$ and
$(\rho = 0, A_{\rm E} = \sqrt{g/2\pi}\,\Lambda)$ corresponding to a
broken chiral symmetry. The first minimum $(\rho = M, A_{\rm E} = 0)$
is deeper than the second one $(\rho = 0, A_{\rm E} =
\sqrt{g/2\pi}\,\Lambda)$ and is, therefore, energetically preferable.

Thus, one can conclude that the evolution of the massless Thirring
model goes through the formation of scalar and pseudoscalar
fermion--antifermion collective excitations with fermion fields
quantized relative to the minimum of the effective potential $V_{\rm
eff}[\rho, A_{\rm E}]$ at $(\rho = M, A_{\rm E} = 0)$.

We have shown that the ground state of the massless Thirring model
quantized around the absolute minimum $(\rho = M, A_{\rm E} = 0)$ is
described by a BCS--type wave function, which is not invariant under
chiral rotations and provides spontaneous breaking of chiral symmetry.
The fermion condensate of Thirring fermions described by this wave
function does not vanish because the BCS--type wave function is not an
eigenfunction of the chirality operator. For eigenfunctions of the
chirality operator the fermion condensate is zero. Nevertheless, such
a choice of the wave function of the ground state does not destroy the
position of the absolute minimum of the effective potential coinciding
with the energy density of the ground state, the dynamical mass $M$.
This means that the chiral symmetry breaking in the ground state
implies an independence of the ground state on the exact value of
chirality.

In polar degrees of freedom the effective quantum field theory of
collective fermion--antifermion excitations with fermion fields
quantized around this minimum is the quantum field theory of a free
massless (pseudo)scalar field $\vartheta(x)$, which can be constructed
without infrared divergences \cite{FI2}. The former can be reached by
removing the collective zero--mode from the observable modes. Indeed,
since the massless Thirring model is well--defined and does not suffer
from infrared divergences (see also \cite{CH67}) a formulation of the
free massless (pseudo)scalar field $\vartheta(x)$, bosonizing the
massless Thirring model in the chirally broken phase, should not also
suffer from an infrared problem. The ground state of a free massless
(pseudo)scalar field $\vartheta(x)$ is described by the bosonized
BCS--type wave function of the ground state of the massless Thirring
model in the chirally broken phase \cite{FI8,FI9}.

In this connection we would like to remind that, according to Wightman
\cite{AW64}, a quantum field theory of a free massless (pseudo)scalar
field $\vartheta(x)$, suffering from infrared divergences, leads to a
violation of Wightman's positive definiteness condition if Wightman's
observables defined on the test functions $h(x)$ from the Schwartz
class ${\cal S}(\mathbb{R}^{\,2})$.  Due to this problem Wightman
\cite{AW64} and then Coleman \cite{SC73} prohibited the existence of
quantum field theories of free massless (pseudo)scalar fields in
1+1--dimensional space--time. In Coleman's formulation such a
suppression was expressed as a non--existence of Goldstone bosons and
spontaneously broken continuous symmetry in quantum field theories
defined in 1+1--dimensional space--time.

It is interesting to remind that Wightman \cite{AW64} admitted a
possibility for the existence of a quantum field theory of a free
massless (pseudo)scalar field in 1+1--dimensional space. He pointed
out that for Wightman's observables defined on the test functions from
the Schwartz class ${\cal S}_0(\mathbb{R}^{\,2}) = \{h(x) \in {\cal
S}(\mathbb{R}^{\,2}); \tilde{h}(0) = 0\}$, where $\tilde{h}(0)$ is the
Fourier transform of $h(x)$ given by
\[
\tilde{h}(0) = \lim_{k \to 0}\int d^2x\,h(x)\, e^{\textstyle\,-ik\cdot
  x} = 0,\] Wightman's positive definiteness condition should be
fulfilled. As a result there are no objections for the existence of
such a quantum field theory of a free massless (pseudo)scalar field.
As we have shown in \cite{FI4} for test functions taken from the
Schwartz class ${\cal S}_0(\mathbb{R}^{\,2}) = \{h(x) \in {\cal
  S}(\mathbb{R}^{\,2}); \tilde{h}(0) = 0\}$ Coleman's theorem tells
nothing about the non--existence of Goldstone bosons and spontaneously
broken continuous symmetry in 1+1--dimensional
space--time\,\footnote{The irrelevance of the
  Mermin--Wagner--Hohenberg theorem \cite{MWH} to the analysis of
  spontaneously broken chiral symmetry in the massless Thirring model
  and a free massless (pseudo)scalar field theory in 1+1--dimensional
  space--time we discussed in details in \cite{FI2}.}. The removal of
the collective zero--mode from the observable vibrational modes of the
free massless (pseudo)scalar field $\vartheta(x)$ corresponds to
Wightman's assumption to formulate the quantum field theory of a free
massless (pseudo)scalar field $\vartheta(x)$ on the Schwartz class of
test functions ${\cal S}_0(\mathbb{R}^2)$.  

We have reformulated this assertion in terms of the generating
functional of Green functions \cite{FI2}. In \cite{FI9} we have shown
that the collective zero--mode is a classical degree of
freedom. Therefore, it cannot be quantized and used as a fluctuating
degree of freedom in the path--integral describing correlation
functions of massless Thirring fermion fields in the bosonized form.
Thus, the use of the free massless (pseudo)scalar field $\vartheta(x)$
with the classical collective zero--mode removed describes well the
bosonized version of the massless Thirring model in the chirally
broken phase.  We would like to mention that the necessity to remove
the collective zero--mode for the non--linear $\sigma$--model in one
and two--dimensional space has been pointed out by Hasenfratz
\cite{PH84}.

\end{document}